\documentclass[preprint,amsmath,amssymb,english,aps,showpacs,showkeys,pr]{revtex4-1}
\usepackage{graphicx}
\usepackage{graphics}
\usepackage{amsmath}
\usepackage{dcolumn}
\usepackage{amssymb}
\usepackage{bm}
\usepackage[latin1]{inputenc}

\begin{document}
\title{Description for rotating  $C_{60}$ fullerenes via G\"odel-type metric}
\author{Everton Cavalcante}
\email{Electronic address: everton@ccea.uepb.edu.br}
\affiliation{Departamento de F\'isica, Universidade Federal da Para\'iba, Caixa Postal 5008, 58051-970, Jo\~ao Pessoa, PB, Brazil;\\ Centro de Ci\^encias
Exatas e Sociais Aplicadas, Universidade Estadual da Para\'{\i}ba, Patos, PB, Brazil}
\author{Josevi Carvalho}
\email{Electronic address: josevi@ccta.ufcg.edu.br}
\affiliation{Unidade Acad\^emica de Tecnologia de Alimentos, Centro de Ci\^encias e Tecnologia Agroalimentar, Universidade Federal de Campina Grande,
Pereiros, Pombal, PB 58840-000, Brazil}
\author{Claudio Furtado}
\email{Electronic address: furtado@fisica.ufpb.br}
\affiliation{Departamento de F\'{\i}sica, Universidade Federal da Para\'{\i}ba, Caixa Postal 5008, 58051-970, Jo\~ao Pessoa, PB, Brazil}

\begin{abstract}
In this contribution a geometric approach to describe a rotating fullerene molecule with Ih symmetry is  developed. We analyze the quantum dynamics of quasiparticles 
in continuum limit considering a description of fullerene in a spherical solution of the G\"odel-type space-time with  a topological defect. As a result, we study 
the molecule in a  rotating frame. Also we combine the well  know non-Abelian monopole approach with this geometric description, including the case of the presence of the external
Aharonov-Bohm flux. The energy levels  and the persistent current for this study are obtained, and we show that  they depend on the geometrical and topological
properties of the fullerene.  Also, we verify recovering of the  well known results for limiting cases.

\end{abstract}


\keywords{fullerene, geometric theory of defect, Dirac equation in curved space}
\pacs{03.65.Pm, 03.65.Ge, 78.30.Na, 73.23.Ra}

\maketitle

\section{Introduction}
 
 The fullerene discovered by Kroto, Curl and Smalley \cite{Kroto} in 1996 is a new allotropic form of carbon with spherical symmetry. This molecule is formed 
by arranging 12 pentagonal rings combined with 20 hexagonal rings of carbon atoms. Thus, the fullerenes are carbon  allotropic forms which
can be described by curved space, that is, the sphere. The geometric description of the fullerene was firstly developed by Gonzales, Guinea and Vozmediano
\cite{Vozmediano1,Vozmediano2}  with use of a tight-binding model. They  showed that the  Fermi surface is reduced to two points, \textbf{K}-points, 
located in the Brillouin zone, and the electrons in this structure  obey the massless Dirac equation for fermions~\cite{CartroNetoGuineaPeres,
VozmedianoKatsnelsonGuinea, Pachos}. The fullerene can be described by a spherical lattice of Ih symmetry, 
where the presence of pentagonal rings  demonstrates the impact of topological defects in this spherical structure. These defects are 
responsible for the presence of pentagons  and are denominated as disclination. The formation of this disclination in a crystalline
structure can be viewed with the called Volterra process~\cite{Volterra}. In this process, disclination can be created by the cut and glue process. 
Disclinations are topological  defects associated with curvature and are characterized by the Frank vector.  A geometric formulatio for this topological
defects was developed by Katanaev and Volovich~\cite{Katanaev}, in this theory,  the continuum description for disclinations and dislocation is given  with use of
curved space with curvature and torsion.

In two-dimensional systems of condensed matter the presence of disclinations  is quite relevant for physical properties of this systems. The study  
of the influence of disclinations in curved structures of graphene employing Dirac equation approach to describe the dynamics of a quasiparticle in this
material, have been relatively well  performed in the Refs.~\cite{Claudio1, KnuteClaudio1,janna,knut, carvalho, knutpra,knutpla,GeusaeClaudio,LammerteCrespi1,LammerteCrespi2}. 
The effect of curvature introduced by disclination in this carbon structure induces an effective gauge field generated by the variation of the local reference frame 
\cite{Claudio1,CartroNetoGuineaPeres, VozmedianoKatsnelsonGuinea,Pachos,LammerteCrespi1,LammerteCrespi2,Birrel}: 
\begin{equation}
\oint \omega_{\mu}dx^{\mu}=-\frac{\pi}{6}\sigma^{3},
\end{equation}
forcing a mixture of the Fermi points ($\textbf{K}_{\pm}$), and generating  a non-Abelian gauge field (or a K-spin flux) as demonstrated by \cite{LammerteCrespi1,LammerteCrespi2},
which compensates the discontinuity of the Bravais labels (A/B):
\begin{equation}
\oint A_{\mu}dx^{\mu}=\frac{\pi}{2}\tau^{2}
\end{equation}
where $\tau^{2}$ is the second Pauli matrix that mixes the $\textbf{K}_{+}$ and $\textbf{K}_{-}$ components of the spinor on the $\textbf{K}_{\pm}$ space. 
In other words, the mixture of Fermi points induces an effective field arising due to a fictitious magnetic monopole in the center of the Ih fullerene by replacing
the fields of 12 disclinations. 

The first model to describe electrons in $C_{60}$ was proposed by Gonzales, Guinea and Vozmediano \cite{Vozmediano1,Vozmediano2}. In this model 
the fullerene is described by {\bf a} non-Abelian magnetic monopole introduced in the centre of a two-dimensional sphere. The field produced by this monopole  represents
the fields of twelve disclinations  presenting in the Ih symmetry of the fullerene molecule. This model establishes the first approach to  study of fullerene using a
Dirac equation to describe low energy electrons in $C_{60}$ molecule. Another description for fullerene 
 in which a gauge field theory is used to describe topological defects,  was used by Kolesnikov and Osipov~\cite{Osipov}. In this model the pentagonal rings in fullerene are described by two gauge fields,  with one of them is used to introduce the elastic properties of disclinations, and the second gauge field, is a non-Abelian gauge field describing the {\bf K}-spin fluxes. 
In recent years,  some studies of fullerene molecules   were employing the continuous model to describe electronic properties of spheroidal fullerenes,  see~\cite{ppo,ppo2,pin}. Recently, two of us have used the geometric
theory of defects to describe fullerene molecule. In this study the well known non-Abelian monopole approach \cite{Vozmediano1,Vozmediano2} was combined with 
the geometric theory of defects~\cite{Katanaev} to obtain the energy levels and the persistent current~\cite{Everton}. Therein, the Dirac equation
for a spherical geometry with topological defects and in the presence of an Aharonov-Bohm flux was solved, and the eigenfunctions and eigenvalues for this problem were exactly determined. 
 
Recently, the study of theoretical models  consideringthe influence of rotation in curved carbon structures has been  carried out. Shen and 
He~\cite{shen}  used  Schr\"odinger equation to study the arising of Aharonov-Carmi effect \cite{ahacarmi} on a rotating  molecules of fullerene. In a recent study, Lima et. 
al. \cite{jonas,jonas1} investigated the effect of rotation in the electronic spectrum fullerene molecules using non-Abelian gauge fields \cite{Vozmediano1,Osipov} 
to describe the defect and {non-inertial effects, including rotation of the reference frame} in the Dirac equation.  The influence of rotation in nanotubes of carbon was also investigated in \cite{brandao}.  
In the present contribution, we investigate the influence of rotation in the spectrum of energy fullerene molecule using a geometric description and 
the well known non-Abelian monopole approach \cite{Vozmediano1,Vozmediano2}. The geometry of spherical rotating body is introduced via a three-dimensional 
spherical G\"odel solution with presence of topological defects.
 
Here, we associate a doublet of spinors interacting with a characteristic curvature of space and with a curvature  accumulated in a pentagonal 
defects (conical singularities) via gauge field ($A_{\mu}$) from the fictitious magnetic monopole in  the centre. In order to put the contents of spherical fullerene under 
rotation, we assume the mapping of a Ih fullerene in a spherical three-dimensional solution ($l^{2}<0$) of a G\"odel-type space-time  with $z=0$  (See Appendix). Also, 
we study the change in spectrum of the $C_{60}$ molecule when when this is crossed by a magnetic flux tube in the direction z .

This paper is 
organized as follows. In Section 2, we give a brief review of the geometric approach for a fullerene molecule, where  its dynamics   is well represented by 
an effective Dirac equation in curved spaces.  We show how to put this content of matter under rotation through a model of continuous in the spherical
Godel-type metric. In Section 3, we consider the presence of a Aharonov-Bohm magnetic flux and discuss how the presence of magnetic flux shifts the spectrum.  Also, we show the emergence of the persistent current, and we compare how the non-inertial characteristics change the spectrum and this persistent current. In Section 4, we 
discuss the results obtained in the  paper.

\section{Geometric approach to a fullerene molecule in a non-inertial frame}\label{section2}
In this section we  present the approach used to describe a fullerene molecule in a non-inertial frame. For this we consider the mapping of a fullerene under 
rotation in a space-time with spherical symmetry descendant of a G\"odel-type solution of the Einstein field equations. In this space-time, the conical singularities
{\bf induce} a non-zero curvature in space. So, we combine two treatments: a non-Abelian monopole approach that is already very well  known
\cite{Gonzalez Guinea Vozmediano 1, Gonzalez Guinea Vozmediano 2} and the geometry theory of defects of Katanaev and Volovich \cite{Katanaev},  based on the
equivalence between three-dimensional gravity with torsion and the theory of defects in solids. Thus, the molecule with conical singularities can be described 
by a Riemann-Cartan geometry. The choice of one spherical G\"odel-type solution is motivated by the  fact that it represents a cosmological solution where the content
material is  rotated. It is interesting to note that the relationship between the geometric approach and the physics of the electronic structure of the
molecule is well described at low energies, around a few tens of eV around the \textbf{K}-points. It is  still useful for elucidating a long-distance physics, 
since we note that the eigenfunctions of the low-energy levels do not oscillate too rapidly at a distance \cite{Kolesnikov Osipov}. Thus, we consider the 
situation in the neighbourhood  of  the defects. In this geometric approach, we consider the fullerene molecule under rotation in terms of a two-dimensional spherical
geometry with a  rotation is labelled  as $\Omega$. Now, we consider G\"odel solution of spherical symmetry, that correspond a rotating body with vorticity (rotation) 
about the $z$-axis (see Appendix \ref{app} for details of G\"odel-type metric solution in Gravitation).

In this solution which was well studied in the context of gravity by \cite{Reboucas Tiomno, Reboucas Aman Teixeira, Galvao Reboucas Teixeira Silva}, the G\"odel metric 
is described by following line element:
\begin{equation}\label{godelsph}
ds^{2}= - \bigg [ dt + \alpha \Omega \frac{sinh^{2} (lr)}{l^{2}}d\phi \bigg ]^{2} + \alpha^{2} \frac{sinh^{2} (2lr)}{4l^{2}} d\phi^{2} + dr^{2} + dz^{2} \mbox{,}
\end{equation}
with the variables ($r,\phi,z,t$) can take, respectively, the following values: $0\le r < \infty$, $0\le \phi \le 2\pi$, $- \infty < (z,t)<\infty$, and $\alpha$ is
related to the angular sector $\lambda$  removed/inserted from/into a spherical sheet in order to form the two conical defects in the sphere
by the expression $\alpha = 1 \pm \lambda / 2\pi$. Indeed, to respect the symmetries of the carbon network, $\lambda$ can only be $\pm N\pi / 3$, where $N$ is
an integer in the interval $(0,6)$. Values of $\alpha$ in the interval, $0<\alpha <1$, mean that we remove a sector of the sphere to form two topological 
defects in the antipodal point. Taking into account the theory  of Katanaev and Volovich \cite{Katanaev}, the elastic continuous medium  with topological 
defect is represented by a three-dimensional space with curvature and torsion. We can see that the spherical G\"odel-type solution as a description of a
spherical body with defects rotating around the $z$-axis. As we are studying the fullerene like a $C_{60}$ buckyball, it is appropriate to do $dz=0$ in 
Eq. (\ref{godelsph}), and we introduce the new convenient coordinates: $R=i/2l$ and $\theta=r/R$, resulting in

\begin{equation}
ds^{2}= - \bigg [ dt + 4 \alpha \Omega R^{2} \sin^{2} \bigg ( \frac{\theta}{2} \bigg ) d\phi \bigg ]^{2}
+ R^{2} \big ( d\theta^{2} + \alpha^{2} \sin^{2} \theta d\phi^{2} \big )
\end{equation}

It is noteworthy that we are adopting natural units $c=\hbar=G=1$. Also, when we consider $\Omega=0$ and $\alpha=1$, the metric reduces to the Minkowski space. 
When $l^{2}=\Omega^{2}/2$ and $\alpha=1$ we  recover the original solution obtained by G\"odel \cite{Godel}. This metric  corresponds to the elastic space-time 
surrounding the defect  and provides all the informations required to characterize the physical system.

The bases of this space-time are known as tetrads (${e^{a}}_{\mu}(x)$), which are defined at each point in space-time by a local reference
frame $g_{\mu \nu}(x)=\eta_{ab}{e^{a}}_{\mu}{e^{b}}_{\nu}$. The tetrad and {\bf its} inverse, ($e^{\mu}_{a}=\eta_{ab}g^{\mu \nu}e^{b}_{\nu}$), 
satisfy the orthogonal relationships: $e^{a}_{\mu}e^{b \mu}=\eta^{ab}$, $e^{a}_{\mu}e^{\mu}_{b}=\delta^{a}_{b}$, $e^{\mu}_{a}e^{a}_{\nu}=\delta^{\mu}_{\nu}$, 
and map the space-time reference frame via the local reference frame \cite{Birrel Davies}:
\begin{equation}
ds^{2}=g_{\mu \nu}dx^{\mu}dx^{\nu}=e^{a}_{\mu}e^{b}_{\nu}\eta_{ab}dx^{\mu}dx^{\nu}=\eta_{ab}\theta^{a}\theta^{b} \mbox{.}
\end{equation}

Greek indices ($\mu, \nu$) run for space-time frame coordinates and the Latin indices ($a,b$) run for local frame coordinates. For the local 
reference frame ($\theta^{a}=e^{a}_{\mu}(x)dx^{\mu}$), we choose the tetrads: 

\begin{equation}
{e^{a}}_{\mu}=
\left( \begin{array}{ccc}
{e^{0}}_{t} & {e^{0}}_{\theta} & {e^{0}}_{\phi} \\
{e^{1}}_{t} & {e^{1}}_{\theta} & {e^{1}}_{\phi} \\
{e^{2}}_{t} & {e^{2}}_{\theta} & {e^{2}}_{\phi} \\
\end{array} \right)=
\left( \begin{array}{ccc}
1 & 0 & 4 \alpha \Omega R^{2} \sin^{2} \big ( \frac{\theta}{2} \big ) \\
0 & R & 0 \\
0 & 0 & \alpha R \sin \theta \\
\end{array} \right) \mbox{.}
\label{tetrada}
\end{equation}

The one-form or spin connections (${{\omega_{\mu}}^{a}}_{b}$) are obtained by the variation of the local reference frame along the a closed curve, $\delta e^{a}_{\mu}$. Namely

\begin{equation}
{{\omega_{\mu}}^{a}}_{b}=-e^{a}_{\beta} \big( \partial_{\mu}e^{\beta}_{b} + \Gamma^{\beta}_{\mu \nu}e^{\nu}_{b} \big ) \mbox{,}
\end{equation} 
where $\Gamma^{\beta}_{\mu \nu}$ are the Christoffel symbols. Another most immediate way to obtain the one-form connection is through the first of the
Maurer-Cartan structure equations:

\begin{equation}
d\theta^{a}+{\omega^{a}}_{b} \wedge \theta^{b}=0 \mbox{.}
\label{Maurer-Cartan equation}
\end{equation}
For the one-forms we get the following connections 
(${\omega^{a}}_{b}={{\omega_{\mu}}^{a}}_{b}dx^{\mu}$):
${{\omega_{\phi}}^{0}}_{1}=-{{\omega_{\phi}}^{1}}_{0}= 2 \alpha \Omega R \sin \theta$, ${{\omega_{\phi}}^{2}}_{1}=-{{\omega_{\phi}}^{1}}_{2}= \alpha \cos \theta$, 
and ${{\omega_{\theta}}^{0}}_{2}=-{{\omega_{\theta}}^{2}}_{0}=  2 \Omega R$. Thus the spinorial connections ($\Gamma_{\mu}(x)=\frac{i}{4}\omega_{\mu ab}\Sigma^{ab}$), 
are described as the components of the doublet related to the \textbf{K}-points as a matrix in ($2+1$)-dimensions
where the Dirac matrices $\gamma^{a}$ are reduced in our case to the Pauli matrices $\gamma^{a}=\sigma^{a}$, and the matrix $\sigma^{0}=I$ is the $2 \times 2$ 
identity matrix, thereby determining the pseudo-spin degrees of freedom. So, the resulting spinorial connections  read as follows: 
\begin{equation}
\left\{ \begin{array}{ll}
\Gamma_{\phi}=\frac{i}{2}\big (\alpha \cos \theta \sigma_{3} - 2 \alpha \Omega R \sin \theta \sigma_{2} \big ) \\
\Gamma_{\theta}=i\Omega R \sigma_{1} \\
\end{array} \right.
\label{spinorial connections}
\end{equation}

Next, we solve the Dirac equation on the surface of a sphere with a fictitious magnetic monopole at its centre. The charge $g$ of the fictitious magnetic 
monopole is adjusted by adding up the individual fluxes of all the lines\cite{Vozmediano2}
\begin{equation}
g=\frac{1}{4\pi}\sum_{i=1}^{N}\frac{\pi}{2}=\frac{N}{8},
\label{monopole charge}
\end{equation}
where $N$ being the number of conical singularities on the surface. Note that for the buckyball $C_{60}$, the structure is that of a truncated icosahedron 
(where $N=12$). Thus we have $g=\frac{3}{2}$, which is compatible with the standard quantization condition of the monopole charge \cite{Gonzalez Guinea Vozmediano 2, Coleman}. 
The spectrum is obtained by solving the covariant Dirac operator,
\begin{equation}
-i \hbar V_{f} \sigma^{a}e^{\mu}_{a}(\nabla_{\mu}-iA_{\mu})\psi=0 \mbox{,} \qquad a=0,1,2 \mbox{,} \quad \mu =t, \theta, \phi \mbox{,}
\label{Dirac operator}
\end{equation}
knowing that $\nabla_{\mu}=\partial_{\mu}-\Gamma_{\mu}$,  and now we assume $V_{f}=1$ is the Fermi velocity. In such way there is a non-Abelian gauge field ($A_{\mu}$) that arises due to the \textbf{K}-spin fluxes.
This 't Hooft-Polyakov monopole must be compatible with the standard quantization condition, and it is well reported by 
\begin{equation}
A_{\phi}=g\cos \theta \tau^{(2)}=\frac{3}{2}\cos \theta \tau^{(2)} \mbox{.}
\label{monopole field}
\end{equation}
Furthermore, it is noteworthy that $\tau^{(2)}$ acts only in the space of $\textbf{K}_{\pm}$ spinor components, while $\sigma^{a}$, existing in spinorial
connection ($\Gamma_{\mu}$), only acts on the geometry. When these matrices operate in different subspaces, we can decouple the doublet ($\psi^{\pm}$). In this case
we will have a rotation in the monopole field, thus we obtain a frame where $\tau^{(2)}$ is diagonal:

\begin{equation}
\oint A_{\mu}^{Rot.}dx^{\mu}=A_{\phi}^{Rot.}=U^{\dagger}A_{\phi}U=A_{\phi}^{k}=
\left\{\begin{array}{ll}
g\cos \theta ,  & \textrm{if $k=(+)$}\\
-g\cos \theta , & \textrm{if $k=(-)$}\\
\end{array} \right.
\label{monopole rotation}
\end{equation}
where

\begin{equation}
U=\frac{1}{\sqrt{2}}
\left( \begin{array}{ccc}
1 & 1 \\
i & -i \\
\end{array} \right)
\end{equation}

Based on a rotation, we can separate the $\textbf{K}_{\pm}$ pseudo-spin components on the doublet. In this case (\ref{Dirac operator}) is reduced to the following

\begin{equation}
-i \hbar  \sigma^{a}e_{a}^{\mu}( \partial_{\mu} - \Gamma_{\mu} - i A_{\mu}^{k} ) 
\psi^{k} = 0
\label{Dirac equation}
\end{equation}

Hence we can carry out our study in the context of eigenvalues problem just defining explicitly the quantum numbers for $\partial_{t}$ and $\partial_{\phi}$. That is, we employ
an \textit{ansatz}: $\psi^{k}(t,\theta,\phi)=\exp(-i \epsilon t)\exp(im\phi)\psi^{k}_{n,m}(\theta)$. However, before that, let us consider the influence of the
Aharonov-Bohm magnetic flux ($\Phi_{B}$) generated by a magnetic string passing through the north pole to the south pole of sphere. 

\section{The rotating Fullerene in the presence of  Aharonov-Bohm flux tube}\label{section3}
In this section, we consider the geometric model described by the Dirac equation (\ref{Dirac equation}) in the presence of Aharonov-Bohm flux. This flux represents 
a magnetic string that goes from pole to pole in the $C_{60}$ buckyball. This is the same as assuming the molecule  to be under the influence of a magnetic field
$\vec{B}=B_{z}\hat{z}$ where $B_{z}=\Phi_{B}\delta(r)$. This flux is associated with a vector potential in the local reference frame by \cite{Everton, Furtado Bezerra Moraes, Carvalho Passos Furtado Moraes}

\begin{equation}
A_{\phi, MS}=\frac{\Phi_{B}}{2\pi}.
\label{vector potential to chiral string}
\end{equation}
Thus, using the minimal coupling 
to include the potential vector (\ref{vector potential to chiral string}) 
in Dirac equation (\ref{Dirac equation}), after replacing the inverse of the tetrads ($e_{a}^{\mu}$) and spinorial connections ($\Gamma_{\mu}(x)$), and considering
$\lambda=\frac{\epsilon R}{\hbar}$, we find
 
\begin{equation}
\bigg [ \frac{d}{d\theta} + \bigg ( \frac{1}{2} + k \frac{g}{\alpha} \bigg ) \cot \theta - \frac{k}{\alpha \sin \theta} \bigg ( m - \frac{\Phi_{B}}{2\pi} + 
4 \alpha \Omega \lambda \sin^{2} \bigg ( \frac{\theta}{2} \bigg ) \bigg ) \bigg ] \psi^{(k)}_{n,m} = i \lambda \psi^{(-k)}_{n,m}
\label{eingenvalues equation}
\end{equation}
Note that we define explicitly the $\phi$-coordinate quantum number ($m=j\pm\frac{1}{2}=0,\pm\frac{1}{2},\pm\frac{3}{2},\dots$) and the stationary character of the problem
(energy spectrum ($\lambda$)) through the \textit{ansatz}: $\psi^{k}(t,\theta,\phi)=\exp(-i \epsilon t)\exp(im\phi)\psi^{k}_{n,m}(\theta)$. Also, when $\Omega=0$, we 
recover the inertial case described in references \cite{Everton, Kolesnikov Osipov, Imura Yoshimura Takane Fukui, Abrikosov} in their respective conditions. These two
components of the doublet ($\psi^{\pm}_{n,m}$) combine to give:

\begin{eqnarray}
&& \bigg \{ \frac{1}{\sin \theta} \frac{d}{d \theta} \sin \theta \frac{d}{d \theta} - \frac{1}{\sin^{2} \theta} \bigg [ \frac{1}{\alpha^{2}} 
\bigg ( m - \frac{\Phi_{B}}{2\pi} \bigg )^{2} - \bigg ( k + \frac{2g}{\alpha} \bigg ) \bigg ( m - \frac{\Phi_{B}}{2\pi} \bigg )\frac{\cos \theta}{\alpha} 
+ \frac{g}{\alpha} \bigg ( \frac{g}{\alpha} + k \bigg ) + \frac{1}{4} \bigg ] +
\nonumber\\ & &
+ \lambda^{2} - \frac{1}{4} + \frac{g^{2}}{\alpha^{2}} - 4 \Omega \lambda \frac{\sin^{2} \theta / 2}{\sin \theta} \bigg [ k + \frac{2}{\alpha} 
\bigg ( m - \frac{\Phi_{B}}{2\pi} - g \cos \theta \bigg ) + 4 \Omega \lambda \sin^{2} \frac{\theta}{2} \bigg ] \bigg \} \psi^{k}_{n,m} = 0
\end{eqnarray}

We will follow the same way as in \cite{Everton}. That is to say that the asymptotic limits do not change when the matter  rotates. Moreover, 
our alternative approach for finding the spectrum $\lambda(\epsilon)$ is based on changing the coordinates $x=\cos^{2}(\theta/2)$ and the \textit{ansatz}

\begin{equation}
\psi^{k}_{n,m}=x^{C_{+}}(1-x)^{C_{-}}H^{k}_{n,m}(x) \mbox{,}
\label{ansatz}
\end{equation}
with

\begin{equation}
C_{\pm}=\frac{1}{2} \bigg | \frac{1}{\alpha} \bigg ( m -\frac{\Phi_{B}}{2\pi} \bigg ) \pm \frac{1}{2} \bigg ( k + \frac{2g}{\alpha} \bigg ) 
\mp \frac{\Omega}{2\alpha} \bigg ( m - \frac{\Phi_{B}}{2\pi} + g \bigg ) \bigg | \mbox{.}
\label{ansatz coefficients}
\end{equation}
Thus, after replacing the \textit{ansatz} (\ref{ansatz}), we obtain an equation  for $H^{k}_{n,m}(x)$:

\begin{eqnarray}
&&\bigg [ x(1-x)\frac{d^{2}}{dx^{2}}+\bigg \{ \bigg | \frac{1}{\alpha} \bigg ( m - \frac{\Phi_{B}}{2\pi} \bigg ) + 
\frac{1}{2}\bigg ( k + \frac{2g}{\alpha} \bigg ) - \frac{\Omega}{2\alpha} \bigg ( m - \frac{\Phi_{B}}{2\pi} + g \bigg ) \bigg | + 1 +
\nonumber\\ & &
- 2 \bigg ( \bigg | \frac{1}{\alpha} \bigg ( m - \frac{\Phi_{B}}{2\pi} \bigg ) \bigg | + 1 \bigg ) x \bigg \} \frac{d}{dx} -
\bigg | \frac{1}{\alpha} \bigg ( m - \frac{\Phi_{B}}{2\pi} \bigg ) \bigg | \bigg ( \bigg | \frac{1}{\alpha} \bigg ( m - \frac{\Phi_{B}}{2\pi} \bigg ) \bigg | + 1 \bigg ) +
\nonumber\\ & &
+ (1-8\Omega^{2})\lambda^{2} - 2\Omega \lambda \bigg [ 1 + \frac{2}{\alpha} \bigg ( m - \frac{\Phi_{B}}{2\pi} + g \bigg ) \bigg ] -
\frac{1}{4}+\frac{g^{2}}{\alpha^{2}} \bigg ] H^{k}_{n,m}(x)=0 \mbox{.}
\label{hiper}
\end{eqnarray}
Note that equation (\ref{hiper}) is very similar  to the standard hypergeometric equation:

\begin{equation}
\bigg [x(1-x)\frac{d^{2}}{dx^{2}}+\big \{\mu + 1 - ( \mu + \nu + 2 ) x \big \} \frac{d}{dx} + n(n + \mu + \nu + 1) \bigg ]F(A,B,C,x)=0
\label{standard hiper}
\end{equation}
Indeed, to ensure the finiteness of the wave function,   the (\ref{hiper}) should behave as a confluent hypergeometric series. Thus, it is possible to compare
(\ref{hiper}) to (\ref{standard hiper}), provided that

\begin{equation}
(1 - 8\Omega^{2})\lambda^{2} - 2 \Omega \bigg [ 1 + \frac{2}{\alpha} \bigg ( m - \frac{\Phi_{B}}{2\pi} + g \bigg ) \bigg ] \lambda -
\bigg ( n + \bigg | \frac{1}{\alpha} \bigg ( m - \frac{\Phi_{B}}{2\pi} \bigg ) \bigg | + \frac{1}{2} \bigg )^{2} + \frac{g^{2}}{\alpha^{2}} = 0 \mbox{.}
\label{eq delta}
\end{equation}
Thereby, we find the spectrum for the particles in model by solving (\ref{eq delta}), resulting in

\begin{eqnarray}
\label{espectro}
&&\epsilon_{n,m} = \frac{\hbar}{2 R (1 - 8\Omega^{2})} \bigg \{ 4\Omega \bigg [ \frac{1}{\alpha} \bigg ( m - \frac{\Phi_{B}}{2\pi} \bigg ) + \frac{1}{2}
+ \frac{g}{\alpha} \bigg ] +
\\ & &
 \pm 2 \sqrt{2 \bigg [ \frac{1}{\alpha} \bigg ( m - \frac{\Phi_{B}}{2\pi} \bigg ) + \frac{1}{2} + \frac{g}{\alpha} \bigg ]^{2} \Omega^{2} -
 (1-8\Omega^{2}) \bigg [ \frac{g^{2}}{\alpha^{2}} - \bigg ( n + \bigg | \frac{1}{\alpha} \bigg (  m - \frac{\Phi_{B}}{2\pi} \bigg ) \bigg | +
 \frac{1}{2} \bigg )^{2} \bigg ]} \bigg \}\nonumber
\end{eqnarray}
It is interesting {\bf to} note that for $\Omega=0$ we restore the inertial case \cite{Everton}. Moreover, it is worth noting that the inversely proportional 
dependence between $\epsilon_{n,m}$ and $R$ do not change the separation between levels. In fact $\Phi_{B}$ operate as a shift in the $z$-component of the
angular momentum. Also {\bf it} should be noted that when $\alpha=1$, $\Phi_{B}=0$ and $\Omega=0$, we recover the results obtained by Kolesnikov and Osipov 
\cite{Kolesnikov Osipov}. Also, when $g=0$ and $\Phi_{B}=0$, we replicate the approach stated by Imura \cite{Imura Yoshimura Takane Fukui}, for a dynamics 
of the fermions in a spherical topological insulator. 
Now, consider the following limit $\Omega <<1$ in (\ref{espectro}) we obtain the following equation:
\begin{eqnarray}
\epsilon_{n,m} = \frac{\hbar }{2 R} \bigg \{ 4\Omega \bigg [ \frac{1}{\alpha} \bigg ( m - \frac{\Phi_{B}}{2\pi} \bigg ) + \frac{1}{2} + \frac{g}{\alpha} \bigg ] 
 \pm 2 \sqrt{  \bigg ( n + \bigg | \frac{1}{\alpha} \bigg (m - \frac{\Phi_{B}}{2\pi } \bigg ) \bigg | + \frac{1}{2} \bigg )^{2}   -\frac{g^{2}}{\alpha^{2}}}\bigg \}.
\label{espectro1}
\end{eqnarray}
 Note that the equation (\ref{espectro1}) is the same found in Ref. \cite{jonas} for $\Phi=0$ and $\alpha=1$. In this way, in the  slow rotation limit  we recover
the results obtained in \cite{jonas}.  Also, the first term is the same contribution obtained by Shen \cite{shen} for Aharonov-Carmi \cite{ahacarmi} effect.

\subsection{The obtaining of the persistent current}
Next, we calculate the persistent current for our model of fullerene drilled by a chiral magnetic string. Indeed, persistent currents were first observed in 
superconducting rings \cite{Buttiker},  while studying the transition around the critical temperature under the influence of an external magnetic field. 
Then, after removing the external field, a residual current is observed. It should be noted that these persistent currents are not caused by outside sources, but 
feature a quantum effect, and  present not only in superconductors but also in usual  conductor and semiconductor  materials. It is calculated using the Byers-Yang relation 
\cite{Byers Yang} given by,
 
\begin{equation}
I=-\sum_{n,m} \frac{\partial \epsilon_{n,m}}{\partial \Phi_{B}} \bigg |_{T=0} \mbox{.}
\label{Byers Yang equation}
\end{equation}
This relation expresses the persistent current for $T=0$ as a derivative of the energy with respect to the magnetic flux of the chiral string. We now obtain the 
following expression for the persistent current,  

\begin{eqnarray}
&&I=\frac{\hbar }{2\pi \alpha R}\sum_{n,m} \bigg \{ 2\Omega +
\nonumber\\ & &
+ \frac{(1-8\Omega^{2})\bigg [ n + \bigg | \frac{1}{\alpha} \bigg ( m - \frac{\Phi_{B}}{2\pi} \bigg ) \bigg | + \frac{1}{2} \bigg ] +
2\Omega^{2} \bigg [ \frac{1}{\alpha} \bigg ( m - \frac{\Phi_{B}}{2\pi} \bigg ) + \frac{1}{2} + \frac{g}{\alpha} \bigg ]}{\sqrt{2\bigg [ \frac{1}{\alpha} 
\bigg ( m - \frac{\Phi_{B}}{2\pi} \bigg ) + \frac{1}{2} + \frac{g}{\alpha} \bigg ]^{2}\Omega^{2}-(1-8\Omega^{2})\bigg [ \frac{g^{2}}{\alpha^{2}} -
\bigg ( n + \bigg | \frac{1}{\alpha} \bigg ( m - \frac{\Phi_{B}}{2\pi} \bigg ) \bigg | + \frac{1}{2} \bigg )^{2} \bigg ]}} \bigg \}\mbox{.}
\label{persistent current}
\end{eqnarray}
Note that the persistent current depends on the parameters $\alpha$ and $g$, which characterize the presence of topological defects, as well  as  the  rotation parameter
$\Omega$. We recover the inertial case \cite{Everton} when $\Omega=0$.  Now, we consider the limit where $\Omega <<1$ in (\ref{persistent current}).  In this
way we obtain the following expression for persistent current,
\begin{eqnarray}
I=\frac{\hbar }{2\pi \alpha R}\sum_{n,m} \bigg \{ 2\Omega 
+ \frac{\bigg [ n + \bigg | \frac{1}{\alpha} \bigg ( m - \frac{\Phi_{B}}{2\pi} \bigg ) \bigg | + \frac{1}{2} \bigg ] }{\sqrt{ \bigg ( n + \bigg | \frac{1}{\alpha}
\bigg ( m - \frac{\Phi_{B}}{2\pi} \bigg ) \bigg | + \frac{1}{2} \bigg )^{2}-\frac{g^{2}}{\alpha^{2}} }}\bigg \} \mbox{.}
\label{persistent current1}
\end{eqnarray}
Notice that in the  slow rotation limit $\Omega <<1$, we can separate two  contributions for persistent current. The first term in Eq. (\ref{persistent current1}), 
directly associated to $\Omega$, and the second term, that is the same found in Ref. \cite{Everton}.

\section{CONCLUSION}\label{section4}

We have investigated a geometric description of the rotating  $C_{60}$ fullerenes within the continuum field theory approach. For this model we map the $C_{60}$ 
fullerenes with Ih symmetry  in a two-dimensional spinning sphere
(about the z-axis) with topological defects, using a Katanaev-Volovich theory of a continuous media \cite{Katanaev}. Specifically we describe the matter contents of the molecule like a homogeneous G\"odel-type metric with spherical symmetry \cite{Godel, Reboucas Tiomno, Reboucas Aman Teixeira, Galvao Reboucas Teixeira Silva}.

 Thus we use the description of the molecule in a non-inertial reference frame, by means of a geometric approach of defects. Thus,in the neighborhood of the defects, a mixture of the $K_{\pm}$ Fermi points  occurs, and a generation of a non-Abelian gauge field, described for a t'Hooft-Polyakov monopole,  takes place \cite{Gonzalez Guinea Vozmediano 1, Gonzalez Guinea Vozmediano 2}.
So, we assume all 12 conical singularities
of the molecule to be described by a G\"odel-type metric.  All this accords with the Osipov-Kolesnikov model \cite{Kolesnikov Osipov}, where eigenfunctions of the 
quasiparticles for small quantum levels do not oscillate too rapidly with distance. Also, we assume that our description works well in causal regions within
the set of solutions $l^{2}<0$ thereby avoiding an imaginary vorticity $\Omega^{2}<0$. 

Finally, our contribution naturally leads to additional terms in energy 
levels of the quasiparticles and current persists observed in the molecule that depends  on rotation $\Omega$. As well as, when compared with the different
approaches about inertial frames \cite{Everton, Kolesnikov Osipov, Imura Yoshimura Takane Fukui, Abrikosov}, the limits are well recovered. 
Also, in the slow rotation limit for the molecule ($\Omega <<1$), we obtain the results found in the references \cite{shen,jonas,jonas1}.
So, we conclude that the results found in this contribution, reveal the non-inertial effects in eigenvalues and eigenfunctions of energy, and persistent currents, for any general values of rotation about $z$-axis for the $C_{60}$ fullerene.

\appendix*\section{G\"odel Solution}\label{app}

It is well known that Einstein's general relativity predicts several interesting phenomena, ranging from the classical time dilatation and contraction of space, to
the curvature of space around heavy objects, such as the sun, a black hole, and most recently, the gravitational waves detected in recent months \cite{Abbott Gravitational Waves}.


In the 80s, three works \cite{Reboucas Tiomno, Reboucas Aman Teixeira, Galvao Reboucas Teixeira Silva} examined in  more detail the problem of causality in the 
Godel-type solutions in Einstein field equations. In all, it is possible to distinguish three different classes of solutions when we study the problem 
in cylindrical coordinates:

\begin{equation}
ds^{2}= - [dt + H(r)d\phi ]^{2} + D^{2}(r)d\phi^{2} + dr^{2} + dz^{2} \mbox{.}
\label{metrica Godel 1}
\end{equation}
Wherein $H(r)=\frac{\Omega}{l^{2}}\sinh^{2} (lr)$, and $D(r)=\frac{1}{2l}\sinh (2lr)$, where  $\Omega$ and $l$ are real constants. Also $\Omega=\frac{H'}{2D}$ 
and $l^{2}=\frac{D''}{4D}$ are necessary conditions \cite{Raychaudhri Thakurta} for the homogeneity of space-time. Also 
\cite{Reboucas Tiomno, Reboucas Aman Teixeira, Galvao Reboucas Teixeira Silva} show that these conditions are not only necessary, but are sufficient for 
homogeneity since there are at least five linearly independent Killing vectors. The presence of CTCs is related to the behaviour of the function:
$G(r)=D^{2}(r)-H^{2}(r)$. So, if $G(r)$ is negative {\bf in} a given limited region, this region will have CTCs. We  have three possibilities: (i) there are 
no CTCs, or $l^{2} \ge \Omega^{2}$, (ii) there is an infinite sequence of alternating causal and non-causal regions, or $l^{2} < 0$, and (iii) there is
only one non-causal region, or $0\le l^{2} < \Omega^{2}$.

One can  also define three classes of solutions in metric as the symmetry of space-time with surfaces of constant curvature: (i) flat solutions, or rotation 
cosmic string, when $l^{2}=0$, (ii) solutions with positive spherical curvature when $l^{2}<0$, (iii) and hyperbolic solutions when $l^{2}>0$. Different aspects
of the G\"odel solutions are discussed also in \cite{Barrow 1, Barrow 2, Clifton Barrow, Gleiser Gurses Karasu Ozgur, Josevi}.

{\bf Acknowledgements}

We thanks to CAPES, CNPQ, CAPES/NANOBIOTEC and FAPESQ for financial support.



\end{document}